\documentclass[preprint2]{aastex6}
\usepackage{epstopdf}

\fullcollaborationName{The Friends of AASTeX Collaboration}

\shorttitle{Jet precession in \object{PG\,1553+113}}
\shortauthors{Caproni et al.}

\begin{document}


\title{Jet precession driven by a supermassive black hole binary system in the BL Lac object \object{PG\,1553+113}}


\author{Anderson Caproni\altaffilmark{1,4}, Zulema Abraham\altaffilmark{2}, Juliana Cristina Motter\altaffilmark{2} and Hektor Monteiro\altaffilmark{3}}




\altaffiltext{1}{N\'ucleo de Astrof\'\i sica Te\'orica, Universidade Cruzeiro do Sul, R. Galv\~ao Bueno 868, Liberdade, S\~ao Paulo, SP, 01506-000, Brazil}
\altaffiltext{2}{Instituto de Astronomia, Geof\'\i sica e Ci\^encias Atmosf\'ericas, Universidade de S\~ao Paulo, R. do Mat\~ao 1226, Cidade Universit\'aria,\\
	S\~ao Paulo, SP, 05508-900, Brazil}
\altaffiltext{3}{Instituto de F\'\i sica e Qu\'\i mica, Universidade Federal de Itajub\'a, Av. BPS 1303-Pinheirinho, Itajub\'a, 37500-903, Brazil}
\altaffiltext{4}{anderson.caproni@cruzeirodosul.edu.br}

\begin{abstract}

The recent discovery of a roughly simultaneous periodic variability in the light curves of the BL Lac object \object{PG\,1553+113} at several electromagnetic bands represents the first case of such odd behavior reported in the literature. Motivated by this, we analyzed 15 GHz interferometric maps of the parsec-scale radio jet of \object{PG\,1553+113} to verify the presence of a possible counterpart of this periodic variability. We used the Cross-Entropy statistical technique to obtain the structural parameters of the Gaussian components present in the radio maps of this source. We kinematically identified seven jet components formed coincidentally with flare-like features seen in the $\gamma$-ray light curve. From the derived jet component positions in the sky plane and their kinematics (ejection epochs, proper motions, and sky position angles), we modeled their temporal changes in terms of a relativistic jet that is steadily precessing in time. Our results indicate a precession period in the observer's reference frame of $2.24\pm0.03$ years, compatible with the periodicity detected in the light curves of \object{PG\,1553+113}. However, the maxima of the jet Doppler boosting factor are systematically delayed relative to the peaks of the main $\gamma$-ray flares. We propose two scenarios that could explain this delay, both based on the existence of a supermassive black hole binary system in \object{PG\,1553+113}. We estimated the characteristics of this putative binary system that also would be responsible for driving the inferred jet precession.  

\end{abstract}




\section{Introduction} \label{sec:intro}

BL Lac objects belong to the blazar class of active galactic nuclei (AGNs) that are characterized by relativistic jets pointing very close to the observer's line of sight, intense high-energy emission, and variability across the entire electromagnetic spectrum.

\begin{figure*}
	\epsscale{1.16}
	\plotone{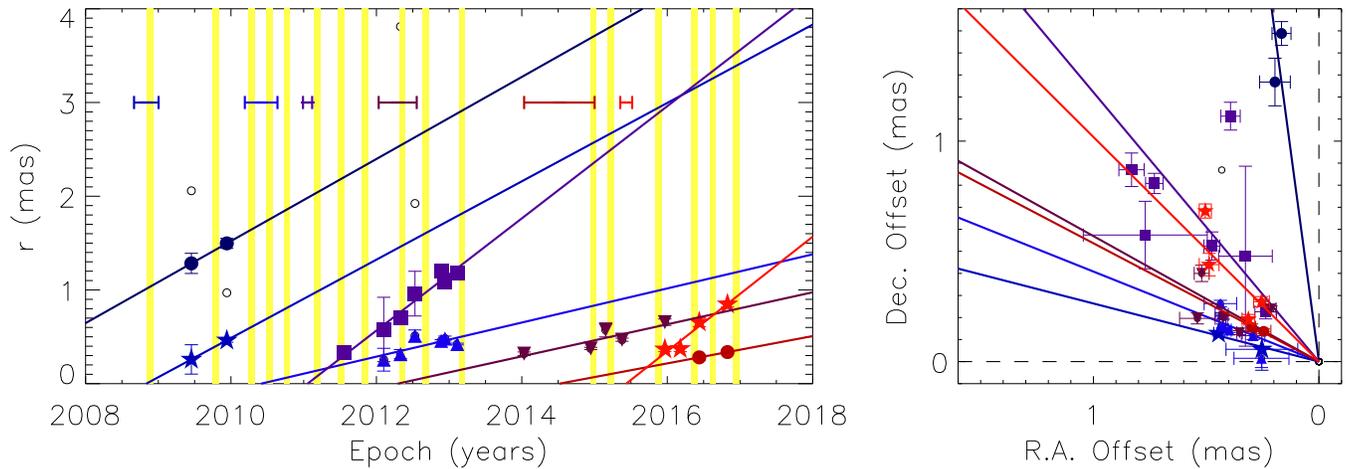}
	\caption{Left panel: Core-component angular separation versus time for the jet components in \object{PG\,1553+113}. Filled symbols represent components identified in this work: navy circles for C1; blue stars for C2; light blue triangles for C3; violet squares for C4; purple inverted triangles for C5; dark red circles for C6; red stars for C7. Unidentified components are shown by open black circles. Continuous lines represent linear regressions applied for the individual jet components, while colored horizontal bars show the 1$\sigma$ uncertainty in their ejection epochs. Vertical yellow bars mark the peak of flare-like features in the public {\it Fermi} $\gamma$-ray light curve (the bar width of 30 days corresponds to the binning used in the referred light curve). Right panel: Right ascension and declination offsets of the same jet components. Straight lines represent their mean position angles. \label{rxt_decxra_plot}}
\end{figure*}

The BL Lac object \object{PG\,1553+113} has an estimated redshift, $z$, of $0.49\pm0.04$\footnote{This value is compatible with the range 0.39-0.58 previously estimated by \citet{dan10}.} \citep{abr15} and it is classified as a high synchrotron peaked source (e.g., \citealt{fan16}). Previous interferometric radio observations of \object{PG\,1553+113} revealed a very compact core--jet structure extending to the northeast direction (e.g., \citealt{tie12,pied14}). This source has been studied in several campaigns that investigated its multi-wavelength and polarimetric variabilities (e.g., \citealt{ack15,rai15,ito16,rai17}), leading to detections of significant rotation in the optical electric vector position angles (EVPAs) that are coincident with $\gamma$-ray flares (e.g., \citealt{bli16,hov16,jer16}).

\object{PG\,1553+113} has gained attention after \citet{ack15} found a periodicity of $2.18\pm 0.08$ years in its $\gamma$-ray emission seen by the \textit{Fermi} Large Area Telescope. Its reliability is strengthened by correlated oscillations in simultaneous radio and optical light curves. According to the authors, possible origins for this periodicity include nearly periodic pulsational accretion flow instabilities that should modulate the energy outflow efficiency, periodic accretion perturbations or jet nutation caused by the presence of a gravitationally bound supermassive black hole (SMBH) binary system, and geometrical models such as jet precession, rotation or helical jet structure in which the observed changes are due to variations in the viewing angle, and, consequently, in the resulting Doppler factor. Such a helical jet model was considered by \citet{rai15} to fit the SEDs of \object{PG\,1553+113} in different brightness states.

A jet periodic precession scenario induced by a binary SMBH system was proposed by \citet{sob17} to explain the  two-year variability. In their model, the jet is carried by the least massive BH in the system, and the observed precession is due to the imprint of the SMBH orbital speed on the jet.

In this work, we identified seven parsec-scale jet components in the source \object{PG\,1553+113} and determined their kinematics, as well as the position angles of their trajectories in the sky plane. The detected variations in their apparent velocities and position angles with time are compatible with precession of a relativistic, parsec-scale jet with a period of $\sim2.24$ years. We also analyzed the feasibility that the jet precession is being caused by a putative binary SMBH system in which the orbital plane of the secondary SMBH is not coincident with the primary accretion disk, inducing torques in its inner parts (e.g., \citealt{abra00,rom00,caab04a,caab04b,cap06,cap13}).

We assume throughout this work $H_0$ = 71 km s$^{-1}$ Mpc$^{-1}$, $\Omega_\mathrm{M}$ = 0.27 and $\Omega_\Lambda$ = 0.73, implying that 1.0 mas = 6.02 pc and 1.0 mas yr$^{-1}$ = 29.24$c$ for \object{PG\,1553+113}, where $c$ is the speed of light.

\section{Observational data and jet kinematics} \label{sec:InterfData}

We used interferometric radio maps of \object{PG\,1553+113} publicly available in the MOJAVE/2 cm Survey Data Archive \citep{lis09}. These 15 GHz data consist of 17 naturally weighted total intensity maps obtained from 2009 to 2016. This interferometric monitoring corresponds to more than 3.3 times the $\gamma$-ray periodicity detected by \citet{ack15}, which makes it suitable for searching for possible signatures of this periodicity in its parsec-scale jet structure. 

We assumed that the individual radio maps of \object{PG\,1553+113} can be decomposed in elliptical Gaussian components. Their structural parameters were determined via the Cross-Entropy (CE) global optimization technique (e.g, \citealt{rubi97,cap09,cap14}). We found the optimal number of Gaussian components in each image following the criteria proposed in \citet{cap14}, leading to a minimum of two and a maximum of five components in the images analyzed in this work. The brightest component in all epochs (flux densities between $\sim$122 and 186 mJy) is the southernmost feature in our model fittings, as well as the closest component to the reference center of the images. It is also unresolved by observations (major axis between $\sim$0.1 and 0.2 mas), and its associated brightness temperature ranged from about $4.3\times10^{10}$ K to $1.5\times10^{11}$ K. We identified this component as the core, where the jet inlet region must be located. 

We found seven jet components (C1--C7) that recede ballistically from the core, as it is shown in Figure \ref{rxt_decxra_plot}. In this figure, it is noted that the ejection epochs of the jet components C2--C7 coincide, within the uncertainties, with the peaks of flare-like features in the public {\it Fermi} $\gamma$-ray light curve (there is no $\gamma$-ray data at the ejection epoch of C1). Since this relation between ejection of new jet components and occurrence of flares in $\gamma$-rays has been found in other AGNs (e.g., \citealt{ott98,jor01,agu11,cut14,lis17}), we believe it provides extra support to our identification scheme given the low number of epochs at which these jet components were detected. Changes in the proper motions ($\sim0.15$--0.61 mas yr$^{-1}$) and in the mean position angles ($\sim 8\degr$--$75\degr$) of the jet components are easily seen in Figure \ref{rxt_decxra_plot}. It suggests that the parsec-scale jet of \object{PG\,1553+113} is not steadily oriented in relation to the observer, as jet precession is a viable scenario to account for such behavior.

\section{A precessing jet in PG\,1553+113} \label{sec:PrecMod}

The precession model used in this work has been successfully applied to other AGNs (e.g., \citealt{abro99,abra00,caab04a,caab04b,cap13}). It assumes a relativistic jet with a constant bulk speed $v$ and an instantaneous viewing angle $\phi$ that varies with time due to precession with a constant precession period measured in the observer's reference frame, $P_\mathrm{prec,obs}$. The relationship between $P_\mathrm{prec,obs}$ and the precession period measured in the source's reference frame, $P_\mathrm{prec,s}$, is calculated from

\begin{equation} \label{Pprecobssource}
P_\mathrm{prec,s} = \frac{P_\mathrm{prec,obs}}{\left(1+z\right)\left(1-\beta\cos\varphi_0\cos\phi_0\right)},
\end{equation}
where $\beta=v/c$, $\varphi_0$ is the semi-aperture angle of the precession cone, and $\phi_0$ is the viewing angle of its axis.

The time modulation in $\phi$ introduced by precession produces a periodic variation in the apparent speed of the jet components, $\beta_\mathrm{obs}$,

\begin{equation} \label{betaobs}
\beta_\mathrm{obs}(\tau_\mathrm{s}) = \frac{\beta\sin\phi(\tau_\mathrm{s})}{\left[1-\beta\cos\phi(\tau_\mathrm{s})\right]},
\end{equation}

\begin{equation} \label{cosphi}
\cos\phi(\tau_\mathrm{s})=-e_\mathrm{x,s}(\tau_\mathrm{s})\sin\phi_0+\cos\varphi_0\cos\phi_0,
\end{equation}
where $\tau_\mathrm{s}=t_\mathrm{s}/P_\mathrm{prec,s}$, $t_\mathrm{s}$ is the time measured in the source's reference frame, as well as in the position angle of the jet components, $\eta$ (positive from north to east),

\begin{equation} \label{eta}
\tan\eta(\tau_\mathrm{s}) = \frac{A(\tau_\mathrm{s})\sin\eta_0+e_{y,\mathrm{s}}(\tau_\mathrm{s})\cos\eta_0}{A(\tau_\mathrm{s})\cos\eta_0-e_{y,\mathrm{s}}(\tau_\mathrm{s})\sin\eta_0},
\end{equation}
and also in the Doppler boosting factor, $\delta$,

\begin{equation} \label{delta}
\delta(\tau_\mathrm{s}) = \gamma^{-1}\left[1-\beta\cos\phi(\tau_\mathrm{s})\right]^{-1},
\end{equation}
where

\begin{equation} \label{A_tau}
A(\tau_\mathrm{s})=e_{x,\mathrm{s}}(\tau_\mathrm{s})\cos\phi_0+\cos\varphi_0\sin\phi_0,
\end{equation}

\begin{equation} \label{exs}
e_\mathrm{x,s}(\tau_\mathrm{s})=\sin\varphi_0\sin(\iota2\pi\Delta\tau_\mathrm{s}),
\end{equation}

\begin{equation} \label{eys}
e_\mathrm{y,s}(\tau_\mathrm{s})=\sin\varphi_0\cos(\iota2\pi\Delta\tau_\mathrm{s}),
\end{equation}
$\Delta\tau_\mathrm{s}=\tau_\mathrm{s}-\tau_\mathrm{0,s}$, $\tau_\mathrm{0,s}=t_\mathrm{0,s}/P_\mathrm{prec,s}$ is the precession phase, $\iota$ is the sense of precession ($\iota=1$ for clockwise sense and $\iota=-1$ for counterclockwise sense; see \citealt{cap09} for further details), $\eta_0$ is the position angle of the precession axis in the sky plane (positive from north to east), and $\gamma$ is the bulk jet Lorentz factor.

The elapsed time measured in the observer's reference frame, $\Delta t_\mathrm{obs}$, is related to that inferred in the source's frame as

\begin{equation} \label{elapsed_time}
\frac{\Delta t_\mathrm{obs}}{P_\mathrm{prec,obs}}=\frac{\int_{0}^{\Delta\tau_\mathrm{s}}\delta^{-1}(\tau)d\tau}{\int_0^1\delta^{-1}(\tau)d\tau}.
\end{equation}

The seven free parameters in our precession model, $P_\mathrm{prec,obs}, \iota, \gamma, \eta_0, \phi_0, \varphi_0$ and $\tau_\mathrm{0,s}$ were determined via the CE technique \citep{cap09,cap13}. For each sense of precession, our CE optimization code minimizes a merit function $S(k)$ at iteration $k$ that is defined as

\begin{equation} \label{merit_function}
S(k) = S_1(k) + S_2(k),
\end{equation}
where

\begin{equation} \label{merit_function1}
S_1(k) = -\ln\left\{\prod\limits_{i=1}^{N_\mathrm{d}}  \frac{\exp\left[-\frac{1}{2}\left(S_{\alpha_i}^2(k)+S_{\delta_i}^2(k)+S_{r_i}^2(k)\right)\right]}{\left(2\pi\right)^{3/2}\sigma_{\Delta\alpha_i}\sigma_{\Delta\delta_i}\sigma_{\Delta r_i}}\right\},
\end{equation}

\begin{equation} \label{merit_function2}
S_2(k) = -\ln\left\{\prod\limits_{i=1}^{N_\mathrm{kin}}  \frac{\exp\left[-\frac{1}{2}\left(S_{\beta_{\mathrm{obs},i}}^2(k)+S_{\eta_i}^2(k)+S_{\beta_{\mathrm{obs,}i}^\prime}^2(k)\right)\right]}{\left(2\pi\right)^{3/2}\sigma_{\beta_{\mathrm{obs},i}}\sigma_{\eta_i}\sigma_{\beta_{\mathrm{obs,}i}^\prime}}\right\}, 
\end{equation}
where $N_\mathrm{d}$ is the number of Gaussian components, and $\sigma_{\Delta\alpha_i}$, $\sigma_{\Delta\delta_i}$, and $\sigma_{\Delta r_i}$ are, respectively, the uncertainties in right ascension, declination, and core-component offsets.  In addition, $S_{\alpha_i}(k)= \sigma_{\Delta\alpha_i}^{-1}\left[\Delta\alpha_i-\Delta\alpha_\mathrm{mod_i}(k)\right]$, $S_{\delta_i}(k)= \sigma_{\Delta\delta_i}^{-1}\left[\Delta\delta_i-\Delta\delta_\mathrm{mod_i}(k)\right]$, $S_{r_i}(k)= \sigma_{\Delta r_i}^{-1}\left[\Delta r_i-\Delta r_{\mathrm{mod}_i}(k)\right]$, $\Delta r_i^2 = \Delta\alpha_i^2+\Delta\delta_i^2$, $\Delta r_{\mathrm{mod}_i}^2 = \Delta\alpha_{\mathrm{mod}_i}^2+\Delta\delta_{\mathrm{mod}_i}^2$, $\Delta\alpha_i$ and $\Delta\delta_i$ are, respectively, the right ascension and declination offsets of the jet component $i$ in relation to the core component, and $\Delta\alpha_\mathrm{mod_i}$ and $\Delta\delta_\mathrm{mod_i}$ are, respectively, the right ascension and declination offsets of the jet component $i$ predicted by the precession model (equations (12) and (13) in \citealt{cap13}).

\begin{deluxetable*}{ccccccccc}
	\tabletypesize{\scriptsize}
	\tablecaption{The Precession-model Parameters Optimized by Our CE technique for Both Clockwise and Counterclockwise Senses of Precession. \label{tab_precmodel}}
	\tablewidth{0pt}
	\tablehead{
		\colhead{$\iota$} & \colhead{$P_\mathrm{prec,obs}$} & \colhead{$P_\mathrm{prec,s}$} & \colhead{$\gamma$} & \colhead{$\eta_0$} &
		\colhead{$\phi_0$} & \colhead{$\varphi_0$} & \colhead{$\tau_\mathrm{0,s}$} & \colhead{$S(k_\mathrm{max})$\tablenotemark{a}} \\
		\colhead{} & \colhead{(year)} & \colhead{(year)} & \colhead{} & \colhead{(deg)} & \colhead{(deg)} & \colhead{(deg)} & \colhead{} & \colhead{} 
	}
	\startdata
	1 & 2.340  $\pm$      0.044   &    152.215  $\pm$  29.381   &      26.437  $\pm$      4.407   &      57.465  $\pm$      3.777   &       5.691  $\pm$      1.047   &       5.559  $\pm$      0.470   &       0.797  $\pm$      0.084   &   912.21\\
	-1 & 2.245  $\pm$      0.011   &    29.634  $\pm$  0.562   &      24.629  $\pm$      3.853   &      48.154  $\pm$      0.885   &      15.254  $\pm$      0.171   &      10.059  $\pm$      0.156   &       0.271  $\pm$      0.004   & 30.54\\
	\enddata
	\tablecomments{The uncertainties in each parameter correspond to the 1$\sigma$ level.}
	\tablenotetext{a}{Merit function at the final iteration ($k_\mathrm{max}=65$).}
\end{deluxetable*}

Furthermore, $N_\mathrm{kin}$ is the number of jet components identified kinetically in this work, $S_{\beta_{\mathrm{obs},i}}(k)= \sigma_{\beta_{\mathrm{obs},i}}^{-1}\left[\beta_{\mathrm{obs},i}-\beta_{\mathrm{obs},i}^\mathrm{mod}(k)\right]$, $S_{\eta_i}(k)= \sigma_{\eta_i}^{-1}\tan\left[\eta_{\mathrm{obs},i}-\eta_{\mathrm{obs},i}^\mathrm{mod}(k)\right]$, $S_{\beta_{\mathrm{obs,}i}^{\prime}}(k)= \sigma_{\beta_{\mathrm{obs,}i}^\prime}^{-1}\left[\beta_{\mathrm{obs},i}^{\prime}-\beta_{\mathrm{obs},i}^\mathrm{mod\prime}(k)\right]$, $\beta_{\mathrm{obs},i}$ and $\eta_{\mathrm{obs},i}$ are, respectively, the apparent speed and position angle of the jet component $i$ in the sky plane, while $\sigma_{\beta_{\mathrm{obs},i}}$ and $\sigma_{\eta_i}$ are their respective uncertainties. Primed and non-primed quantities are given in terms of time (in the observer's reference frame) and position angle, respectively, with $\sigma_{\beta_{\mathrm{obs,}i}^\prime} = \sqrt{\sigma_{\beta_{\mathrm{obs},i}}\sigma_{\eta_i}}$.

Following the procedures adopted in \citet{cap14}\footnote{In a few words, it is based on a plot that relates the distribution of the apparent velocities of the jet components and the Doppler boosting factors derived from the core brightness temperatures (e.g., Figure 7 in \citealt{cap14}).}, we estimated the probable range for the bulk Lorentz factor of the parsec-scale jet, roughly between 9.9 and 52.0, and its viewing angle, between 4\fdg9 and 20\fdg5. They are compatible with previous estimates found in the literature (e.g., \citealt{cegu08,rai17}), and they were used as input in our CE precession-model optimizations. The optimal precession period in the observer's reference frame was searched between 1.9 years (the $\gamma$-ray periodicity minus its uncertainty; \citealt{ack15}) and 7.5 years (the approximated interval of the radio interferometric observations). The optimal position angle of the precession axis in the sky plane was searched between 30\degr and 60\degr, which corresponds, respectively, to the mean position angle of the individual jet components subtracted and added by the respective standard deviation.

\begin{figure*}
	\epsscale{1.15}
	\plotone{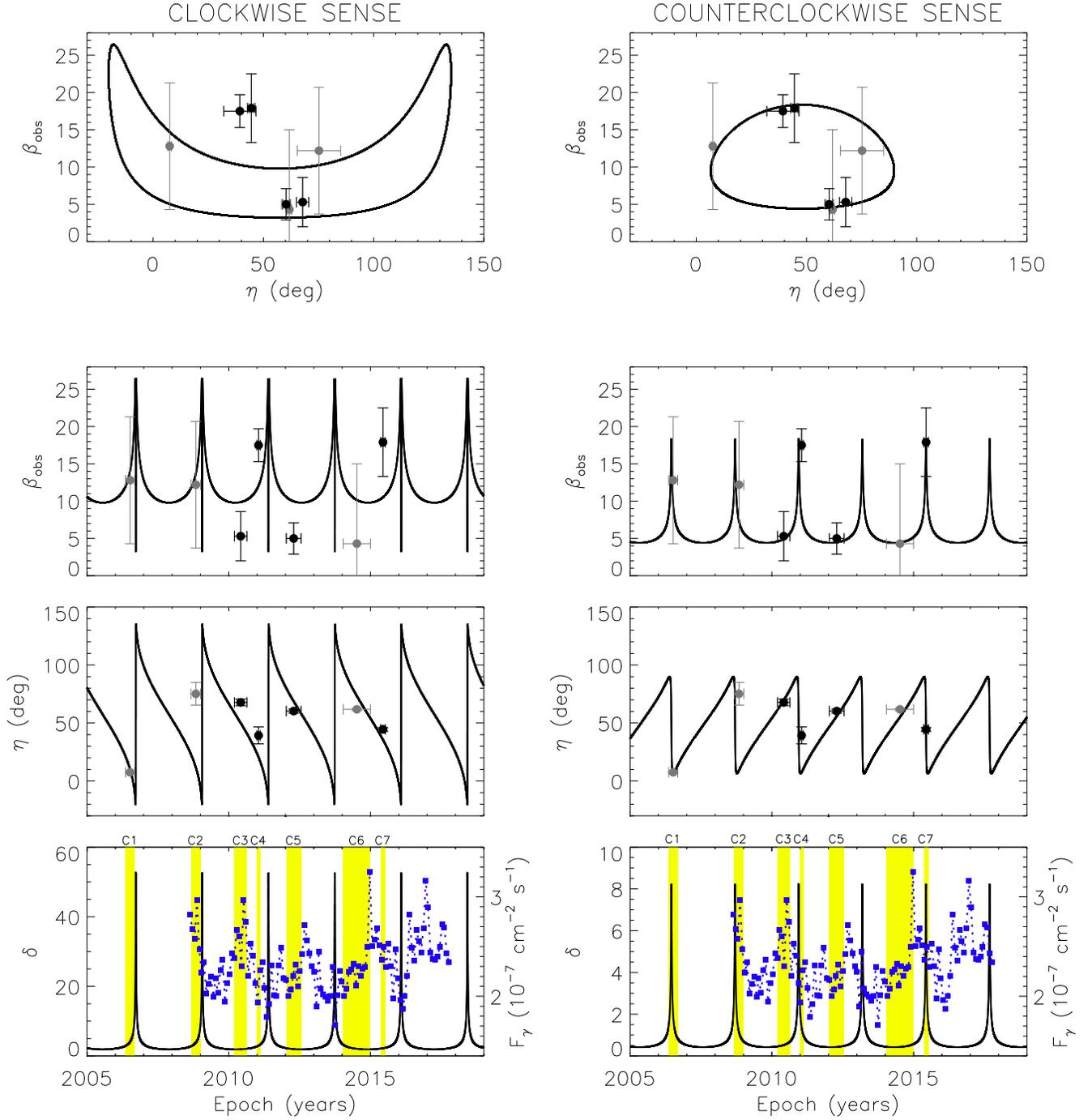}
	\caption{Our CE optimized models for clockwise (left panels) and counterclockwise (right panels) senses of precession of the parsec-scale jet of \object{PG\,1553+113}. Solid curves represent the predictions from the precession models presented in Table \ref{tab_precmodel} in the time-independent $\beta_\mathrm{obs}-\eta$ plane (top panels), as well as these two quantities as a function of time. Circles correspond to the apparent speeds and position angles in the sky plane of the jet components identified in this work. Black color codes jet components with at least three epochs of detection, while gray color refers to those with only two epochs of observation. The bottom panels show the time behavior of the Doppler boosting factor superposed to the the public {\it Fermi} $\gamma$-ray light curve (blue squares). Vertical yellow bars show the 1$\sigma$-uncertainty range for the ejection epochs of the jet components. \label{precmodel_plot}}
\end{figure*}

We show the optimal precession-model parameters for both clockwise and counterclockwise senses of precession in Table \ref{tab_precmodel}. A comparison between our precession models and the kinematic data of the identified jet components is presented in Figure \ref{precmodel_plot}. A visual inspection of this figure indicates that only the counterclockwise model fits the data set, which is corroborated by the values of $S(k=k_\mathrm{max})$ shown in Table \ref{tab_precmodel}.

The counterclockwise jet precession period in the observer's reference frame is compatible with the $\gamma$-ray periodicity reported by \citet{ack15}. However, major $\gamma$-ray flares seem to lead temporal variations in the counterclockwise Doppler factor in Figure \ref{precmodel_plot}. The application of the discrete discrete correlation function (\citealt{edkr88}) confirms it, resulting in a delay of $\sim0.45$ years, suggesting that these $\gamma$-ray flares are not driven by jet precession. We propose two possible explanations for such a delay, both involving the presence of an SMBH binary system in the nucleus of \object{PG\,1553+113}. 

The $\gamma$-ray periodicity could be related to the crossing of the secondary SMBH through the accretion disk of the primary SMBH, as in the BL Lac object \object{OJ\,287} (e.g., \citealt{leva96}). The crossing could induce perturbations in the primary's disk, enhancements of the accretion rate that result in an increase of the jet flow, the formation of subsequent shocks, and the corresponding flares \citep{val00}.

Another possibility is that $\gamma$-ray periodicity is a consequence of inverse Compton scattering of seed photons from the accretion disk of the secondary SMBH by the relativistic particles in the precessing jet. The relative orientation between jet and secondary disk would be modulated by the jet precession, which could introduce a variability in the $\gamma$-ray light curve with a similar period of the jet precession. Interestingly, the jet precession period inferred in this work agrees with the reported $\gamma$-ray periodicity at a $1\sigma$ level.

\section{An SMBH binary system in PG\,1553+113} \label{sec:SMBHBS}

In this section, we estimate the characteristics of a putative SMBH binary system in \object{PG\,1553+113}, assuming that the inferred jet precession is driven by a secondary SMBH with mass $M_\mathrm{s}$ in a non-coplanar circular orbit of radius $R_\mathrm{ps}$ around the primary SMBH with mass $M_\mathrm{p}$ (e.g., \citealt{sil88,katz97,abra00,rom00,caab04a,caab04b,cap06,cap13,rol13}). The parsec-scale jet of \object{PG\,1553+113} is assumed to be associated with the primary SMBH.

Following the formalism employed by \citet{bat00}, and also assuming that the Roche lobe radius can be used as a proxy for the outer radius of the primary's disk (as in \citealt{cap06}), we established a relationship between $P_\mathrm{prec,s}$ and the orbital period of the secondary SMBH, $P_\mathrm{ps,s}$

\begin{equation} \label{Pprec_div_P_ps}
\frac{P_\mathrm{ps,s}}{P_\mathrm{prec,s}}=\left[K(s)\cos\varphi_0\right]\left[\frac{q\left[0.88 f(q)\right]^{3/2}}{\left(1+q\right)^{1/2}}\right] ,
\end{equation}
where $K(s)\approx0.75(s+5/2)/(s+4)$, $s$ is the power-law index of the surface density, $\Sigma$, of the primary's disk ($\Sigma(r)\propto r^s$), $q=M_\mathrm{s}/M_\mathrm{p}$, and $f(q)=\left(0.49q^{2/3}\right)\left[0.6q^{2/3}+\ln{\left(1+q^{1/3}\right)}\right]^{-1}$.

Independent of the values of $s$ and $q$, the stability of the binary system against gravitational-wave losses must be fulfilled to produce a time-steady jet precession. Following \citet{shte83}, the gravitational-wave timescale in the source's reference frame, $\tau_\mathrm{GW,s}$, can be written in terms of $P_\mathrm{prec,s}$ as follows:

\begin{equation} \label{tauGWs_div_Pprecs}
\frac{\tau_\mathrm{GW,s}}{P_\mathrm{prec,s}} = \frac{5c^5K(s)\cos\varphi_0}{\left(128\pi^3\right)^2}\left(\frac{P_\mathrm{ps,s}}{R_\mathrm{ps}}\right)^5\left[0.88 \left(1+q\right)f(q)\right]^{3/2}.
\end{equation}

Setting $\tau_\mathrm{GW,s}\ga 10^5$ years ($\sim 3375P_\mathrm{prec,s}$) in equation (\ref{tauGWs_div_Pprecs}),\footnote{Even though $10^5$ years is an arbitrary choice, it is enough to provide an orbital stability of the secondary against gravitational losses for a long period of time (e.g., after $10^4$ years from now, the orbital radius will shrink to $\sim0.974R_\mathrm{ps}$, decreasing the orbital period and the precession period by a factor of about 0.961 in relation to the present values).} $q$ necessarily must be larger than $\sim0.35$ for $s=0$, increasing to 0.81 for $s=-2$. These limits in $q$ translate to lower limits for $P_\mathrm{ps,s}$ and $R_\mathrm{ps}$: 0.54 years and $1.01\times 10^{-3}$ pc for $s=0$, and 0.59 years and $1.08\times 10^{-3}$ pc for $s=-2$.

It is also possible to estimate a lower limit for the ratio of semi-thickness of the primary accretion disk to radius, $H/r$ in terms of $q$. Following \citet{bat00}, we assumed that any perturbation in the disk propagates in a wave-like regime, which is expected whenever $H/r\ga\alpha$ \citep{pali95}. Thus, it is necessary that 

\begin{equation} \label{wp_div_Omegad_H_div_R}
\left(\frac{H}{R}\right) \gtrsim K(s)\cos\varphi_0q\left[0.88 f(q)\right]^{3}
\end{equation}
for the primary's disk to precess rigidly. For $s=0$, $H/r\gtrsim 0.0027$ if $q\approx 0.35$, while for $s=-2$, $H/r\ga 0.0048$ if $q=0.81$. These lower limits are compatible with geometrically thin and thick accretion disk models available in the literature (e.g., \citealt{shsu73,prin81,abr88,nayi94,nara03}).

It is important to emphasize that all of these estimates do not exclude the possibility that $q>1$, as proposed by \citet{sob17}, as they are in agreement with the two scenarios presented at the end of Section \ref{sec:PrecMod}.

\section{Final remarks} \label{sec:concl}

In this work, we analyzed the viability of the 2.18 year quasi-periodic variability detected in the $\gamma$-ray light curves of the BL Lac object \object{PG\,1553+113} by \citet{ack15} to be associated with the precession of the parsec-scale jet of this source. With this aim, we collected 17 interferometric maps of \object{PG\,1553+113} at 15 GHz, starting in mid 2009 and finishing at the end of 2016 ($\sim$7.5 years of monitoring).

As in \citet{cap14}, we assumed that jet emission can be modeled by bidimensional, elliptical Gaussian sources. We applied our CE global optimization technique to determine the structural parameters of these Gaussian components, being the brightest and the southernmost component identified as the core, where the jet inlet region must be located.

From the derived jet component offsets in right ascension and declination, we identified seven distinct jet components with ballistic trajectories. Their ejection roughly coincided with the occurrence of $\gamma$-ray like flares. In addition, their apparent speeds range from 5.0$c$ to 17.9$c$, while their mean position angles in the sky plane are between 7\fdg5 and 75\fdg2. We interpreted these changes in $\beta_{\mathrm{obs}}$ and $\eta$ as a signature of jet precession.

We used our analytical precession model to fit simultaneously the right ascension and declination offsets of each jet component, as well as their respective apparent velocities and position angles in the sky plane. The best set of precession-model parameters was estimated from the application of the CE global optimization technique in the minimization of the merit function (equation \ref{merit_function}). The counterclockwise sense of precession is strongly favored by the value of the merit function.

Our results indicate that the parsec-scale jet of \object{PG\,1553+113} is highly relativistic ($\gamma\sim 25$), pointing close to the line of sight (roughly between 5\degr and 20\degr), and precessing with a period of $\sim 2.2$ years ($\sim 29.6$ years in the source's reference frame) and a precession angle of $\sim$ 10\degr. Thus, precession introduces a periodic oscillation in the jet viewing angle (roughly between 5\fdg2 and 25\fdg3) that will induce periodic variations in the Doppler boosting factor (between $\sim$ 0.4 and 8.5), and consequently in the observed flux density of \object{PG\,1553+113}. Even though the jet precession period inferred in this work agrees with the reported $\gamma$-ray periodicity, there is a delay of $\sim0.45$ years between the maxima of jet Doppler boosting factor and the peak of the main $\gamma$-ray flares. It is interesting to note that although the value of $\delta$ increases by a factor of $\sim20$ during the jet closest approach to our line of sight, neither radio, optical, or high-frequency light curves presented strong peaks at those times, as would be expected from beaming. This can be explained considering that the total flux is the combined emission of material ejected at different epochs, with different angles relative to the line of sight, so that the beamed contribution must represent a small part of the total flux. We propose two scenarios involving a putative SMBH binary system in \object{PG\,1553+113} that could reconcile jet precession and the periodic occurrence of the main $\gamma$-ray flares (see Section \ref{sec:PrecMod}). Future monitoring of the source's activity in all wavelengths, including radio interferometric studies, should be carried out in order to confirm our results and propositions.

Finally, we estimated the characteristics of the possible SMBH binary system in \object{PG\,1553+113}, assuming that the inferred jet precession is driven by a secondary SMBH in a non-coplanar, circular orbit around the primary SMBH. Adopting the formalism in \citet{bat00}, and imposing stability against gravitational-wave losses, we could estimate a lower limit for $q$: 0.35 for a power-law primary accretion disk with a constant surface density ($s=0$), increasing to about 0.81 for a disk with $s=-2$. These limits in $q$ translates to lower limits for $P_\mathrm{ps,s}$ and $R_\mathrm{ps}$: 0.54 years and $1.01\times 10^{-3}$ pc for $s=0$, and 0.59 years and $1.08\times 10^{-3}$ pc for $s=-2$. Assuming that the primary's disk precesses rigidly and any perturbation propagates wavily through the disk, we could set lower limits for the aspect ratio of the accretion disk in terms of $q$ and $s$: $H/r\ga 0.0027$ for $s=0$ and $q=0.35$, and $H/r\ga 0.0048$ for $s=-2$ and $q=0.81$.

\acknowledgments

A.C. thanks CNPq (grant 305990/2015-2) and FAPESP. J.C.M. thanks CNPq (grant 142041/2013-0). Z.A. acknowledges FAPESP (grant 2015/50360-9). We also thank the anonymous referee for the useful report. This research has made use of data from the MOJAVE database that is maintained by the MOJAVE team (Lister et al., 2009, AJ, 137, 3718).


\begin{thebibliography}{}

\bibitem[Abraham \& Romero(1999)]{abro99} Abraham, Z., Romero, G. E. 1999, \aap, 344, 61

\bibitem[Abraham(2000)]{abra00} Abraham, Z. 2000, \aap, 355, 915

\bibitem[Abramowicz et al.(1988)]{abr88} Abramowicz M. A., Czerny B., Lasota J. P., Szuszkiewicz E. 1988, \apj, 332, 646

\bibitem[Abramowski et al.(2015)]{abr15} Abramowski, A., Aharonian, F., Ait Benkhali, F. et al. 2015, \apj, 802, 65

\bibitem[Ackermann et al.(2015)]{ack15} Ackermann, M., Ajello, M., Albert, A. et al. 2015, \apjl, 813, L41

\bibitem[Agudo et al.(2011)]{agu11} Agudo, I., Marscher, A. P., Jorstad, S. G. et al. 2011, \apjl, 735, L10

\bibitem[Bate et al.(2000)]{bat00} Bate, M. R., Bonnell, I. A., Clarke, C. J., Lubow, S. H., Ogilvie, G. I., Pringle, J. E., Tout, C. A. 2000, \mnras, 317, 773

\bibitem[Blinov et al.(2016)]{bli16} Blinov, D., Pavlidou, V., Papadakis, I. E., et al. 2016, \mnras, 457, 2252

\bibitem[Caproni \& Abraham(2004a)]{caab04a} Caproni, A., Abraham, Z. 2004a, \apj, 602, 625

\bibitem[Caproni \& Abraham(2004b)]{caab04b} Caproni, A., Abraham, Z. 2004b, \mnras, 349, 1218

\bibitem[Caproni et al.(2006)]{cap06} Caproni, A., Livio, M., Abraham, Z., Mosquera Cuesta, H. J. 2006, \apj, 653, 112

\bibitem[Caproni, Monteiro \& Abraham(2009)]{cap09} Caproni, A., Monteiro, H., Abraham, Z. 2009, \mnras, 399, 1415

\bibitem[Caproni et al.(2013)]{cap13} Caproni, A., Abraham, Z., Monteiro, H. 2013, \mnras, 428, 280

\bibitem[Caproni et al.(2014)]{cap14} Caproni, A., Tosta e Melo, Abraham, Z., Monteiro, H., Roland, J. 2014, \mnras, 441, 187

\bibitem[Celotti \& Ghisellini(2008)]{cegu08} Celotti, A., Ghisellini, G. 2008, \mnras, 385, 283

\bibitem[Cutini et al.(2014)]{cut14} Cutini, S., Ciprini, S., Orienti, M. et al. 2014, \mnras, 445, 4316

\bibitem[Danforth et al.(2010)]{dan10} Danforth, C. W., Keeney, B. A., Stocke, J. T., Shull, J. M., Yao, Y. 2010, \apj, 720, 976

\bibitem[Edelson \&  Krolik(1988)]{edkr88} Edelson, R., Krolik, J. 1988, \apj, 333, 646

\bibitem[Fan et al.(2016)]{fan16} Fan, J. H., Yang, J. H., Liu, Y. et al. 2016, \apjs, 226, 20

\bibitem[Hovatta et al.(2016)]{hov16} Hovatta, T., Lindfors, E., Blinov, D. et al. 2016, \aap, 596, A78

\bibitem[Itoh et al.(2016)]{ito16} Itoh, R., Nalewajko, K., Fukazawa, Y. et al. 2016, \mnras, 833, 77

\bibitem[Jermak et al.(2016)]{jer16} Jermak, H., Steele, I. A., Lindfors, E. et al. 2016, \mnras, 462, 4267

\bibitem[Katz(1997)]{katz97} Katz, J. I. 1997, \apj, 478, 527

\bibitem[Lehto \& Valtonen(1996)]{leva96} Lehto, H. J. \& Valtonen, M. J. 1996, \apj, 460, 207

\bibitem[Lister et al.(2009)]{lis09} Lister, M. L., Aller, H. D., Aller, M. F. et al. 2009, \aj, 137, 3718

\bibitem[Lisakov et al.(2017)]{lis17} Lisakov, M. M., Kovalev, Y. Y., Savolainen, T., Hovatta, T., Kutkin, A. M. 2017, \mnras, 468, 4478

\bibitem[Jorstad et al.(2001)]{jor01} Jorstad, S. G., Marscher, A. P., Mattox, J. R., Aller, M. F., Aller, H. D., Wehrle, A. E., Bloom, S. D. 2001, \apj, 556, 738

\bibitem[Narayan \& Yi(1994)]{nayi94} Narayan, R., Yi, I. 1994, \apjl, 428, 13

\bibitem[Narayan(2003)]{nara03} Narayan, R. 2003,\pasj, 55, L69

\bibitem[Otterbein et al.(1998)]{ott98} Otterbein, K., Krichbaum, T. P., Kraus, A., Lobanov, A. P., Witzel, A., Wagner, S. J., Zensus, J. A. 1998, \aap, 334, 489

\bibitem[Papaloizou \& Lin(1995)]{pali95} Papaloizou J. C. B., Lin D. N. C. 1995, \mnras, 438, 841

\bibitem[Piner \& Edwards(2014)]{pied14} Piner, B. G., Edwards, P. G. 2014, \apj, 797, 25

\bibitem[Pringle(1981)]{prin81} Pringle, J. E. 1981, \araa, 19, 137

\bibitem[Raiteri et al.(2015)]{rai15} Raiteri, C. M., Stamerra, A., Villata, M. et al. 2015, \mnras, 454, 353

\bibitem[Raiteri et al.(2017)]{rai17} Raiteri, C. M., Nicastro, F., Stamerra, A. et al. 2017, \mnras, 466, 3762

\bibitem[Roland et al.(2013)]{rol13} Roland J., Britzen S., Caproni A., Fromm C., Gl\"uck C., Zensus A., 2013, \aap, 557, 85

\bibitem[Romero et al.(2000)]{rom00} Romero, G. E., Chajet, L., Abraham, Z., Fan, J. H. 2000, \aap, 360, 57

\bibitem[Rubinstein (1997)]{rubi97} Rubinstein, R. Y. 1997, European Journal of Operational Research, 99, 89

\bibitem[Shakura \& Sunyaev(1973)]{shsu73} Shakura N. I., Sunyaev R. A. 1973, \aap, 24, 337

\bibitem[Shapiro \& Teukolsky(1983)]{shte83} Shapiro, S. L., \& Teukolsky, S. A. 1983, in Black Holes, White Dwarfs, and Neutron Stars (New York: Wiley)

\bibitem[Sillanp\"a\"a et al.(1988)]{sil88} Sillanp\"a\"a, A., Haarala, S., Valtonen, M. J., Sundelius, B., Byrd, G. G. 1988, \apj, 325, 628

\bibitem[Sobacchi, Sormani \& Stamerra(2017)]{sob17} Sobacchi, E., Sormani, M. C., Stamerra, A. 2017, \mnras, 465, 161

\bibitem[Tiet, Piner \& Edwards(2012)]{tie12} Tiet, V. C., Piner, B. G., Edwards, P. G. 2012, Fermi, \& Jansky Proceedings [arXiv:1205.2399]

\bibitem[Valtaoja et al.(2000)]{val00} Valtaoja, E., Ter\"asranta, H., Tornikoski, M., Sillanp\"a\"a, A., Aller, M. F., Aller, H. D., Hughes, P. A. 2000, \apj, 531, 744










\end{thebibliography}
\end{document}